\documentclass[a4paper, 11pt]{article}

\usepackage{a4wide}

\usepackage[english]{babel}
\usepackage[utf8]{inputenc}
\usepackage[T1]{fontenc}

\usepackage{amsthm}

\usepackage[ttscale=.875]{libertine}
\usepackage[libertine]{newtxmath}
\usepackage[scaled=1]{inconsolata}

\usepackage{graphicx}

\theoremstyle{remark}
\newtheorem*{remark}{Remark}

\usepackage{booktabs}

\usepackage[colorlinks=false, hidelinks]{hyperref}

\usepackage{siunitx}
\usepackage{ragged2e}
\newcolumntype{R}[1]{>{\RaggedRight\arraybackslash}p{#1}}

\title{Bad Crypto: Chessography and Weak Randomness of Chess Games}
\author{
  Martin Stanek \\[2ex]
  Department of Computer Science \\
  Faculty of Mathematics, Physics and Informatics \\
  Comenius University \\
  \textsl{martin.stanek@fmph.uniba.sk}
}

\date{}

\begin{document}
\maketitle

\begin{abstract}
\noindent This short communication shows that the Chessography encryption scheme is incorrect, redundant, and the the security claims based on the complexity of chess games are unjustified. It also demonstrates an insufficient randomness in the final chess game positions, which could be of separate interest. \\[2ex]
\end{abstract}

\section{Failures of Chessography}

Chess is an interesting and complex game with a vast number of possible positions. This sometimes leads to ideas for crossing chess with cryptography \cite{A16, K17, S24, S16}. Chessography \cite{K17} is a symmetric encryption scheme. The main idea of this scheme is to use a chess game to encrypt a plaintext block of 32 characters. Plaintext characters are placed on squares where white and black pieces are positioned at the start of the game. These characters are then transformed using the first key, and the game is used to move them on the board. The ciphertext consists of the final position of the pieces on the board, together with additional information that allows to reconstruct the initial positions of the pieces, including those that were captured during the game. The main objections to the quality and strength of the Chessography scheme are summarized in the following paragraphs.

\subsubsection*{Imprecise description of the scheme}

The description of the encryption and decryption algorithms is rather vague. It lacks mathematical formulas, reference implementation, or pseudo-code. It is unclear  whether distinct chess games are used for subsequent plaintext blocks or if a single game is used for all blocks. The provided example also leaves numerous questions unanswered, e.g., what is the exact procedure to produce the final ciphertext, since the text description does not correspond to the ciphertext presented in figures.

\subsubsection*{The scheme is incorrect}

\begin{figure}[ht]
\centering
\includegraphics[scale=0.5]{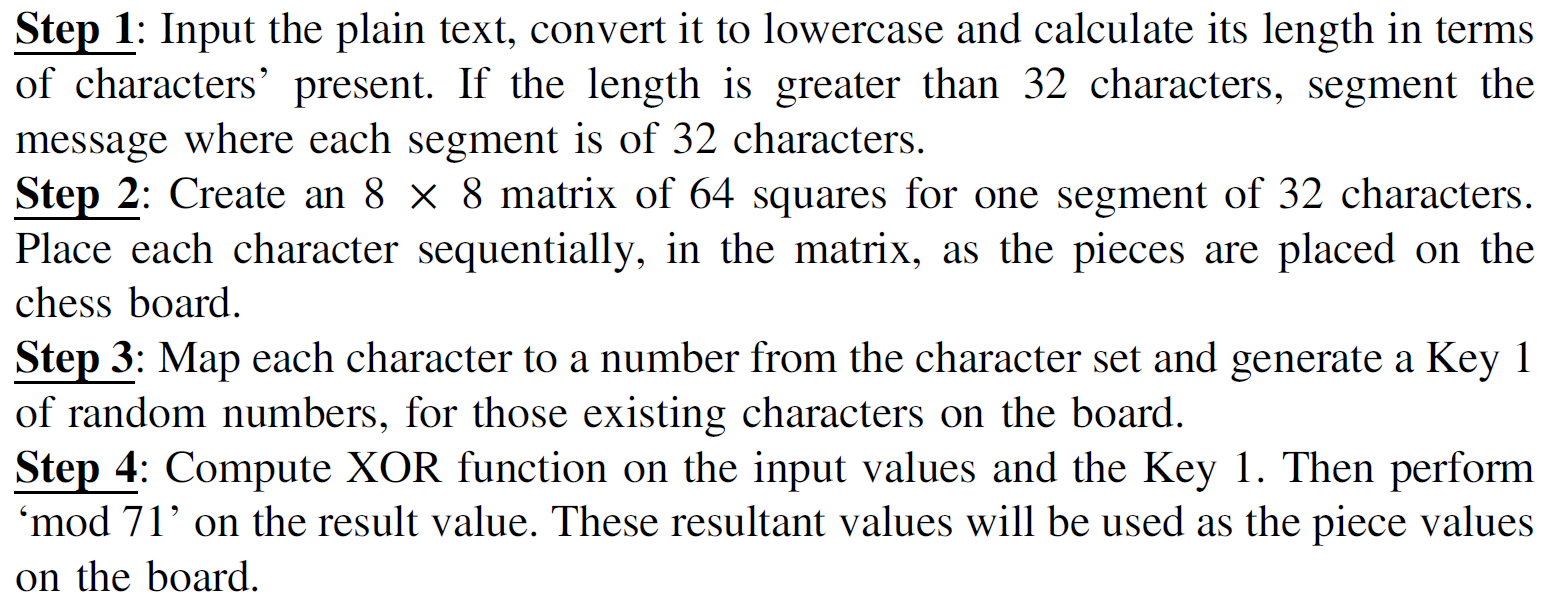}
\caption{First four steps of encryption \cite[Sect. 3.1]{K17}}
\label{fig1-xor}
\end{figure}

The scheme uses an alphabet with $71$ characters, which are encoded as numbers ranging from $1$ to $71$. Figure \ref{fig1-xor} illustrates the initial four steps in the encryption algorithm. There are two notable issues with step 4. A minor issue is that after a modulo $71$ operation, the range changes to $0,\ldots, 70$. 

\begin{figure}[ht]
\centering
\includegraphics[scale=0.5]{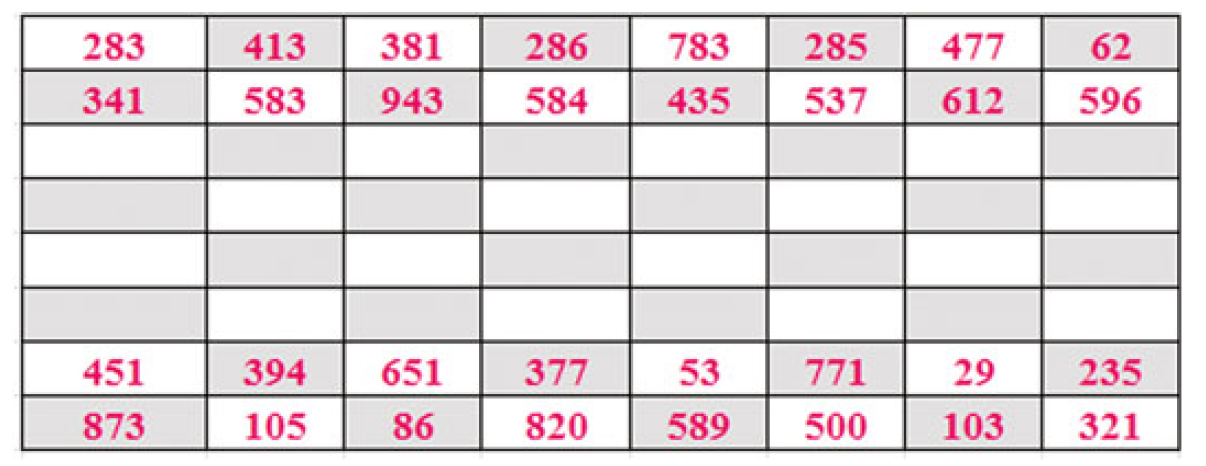}
\caption{Example of key 1 \cite[Fig. 9]{K17}}
\label{fig2-key}
\end{figure}

A major issue is that XOR-ing with randomly chosen ``Key 1'', whose values can be quite large, see Figure \ref{fig2-key}, and performing modulo $71$ operation is not reversible. For instance, consider the plaintext characters encoded as numbers $9$ and $64$, and the key value $62$ for both numbers: 
  \begin{align*}
  (9 \text{ XOR } 62) \bmod 71 =  55 \bmod 71 = 55;\\
  (64 \text{ XOR } 62) \bmod 71 =  126 \bmod 71 = 55.
  \end{align*}

\noindent It is impossible to tell what the original plaintext number was just from the result $55$ and the key value $62$. Hence the step 4 is not reversible. It does not matter what the next transformations are, decryption will not be able to produce the correct plaintext.

\subsubsection*{Chess-related permutation is (sometimes) irrelevant}

Let's assume that the ``XOR-mod'' step is correct, e.g., the scheme uses an alphabet with $64$ characters, numbered from $0$ to $63$, and values in Key 1 are chosen as 6-bit integers. If only a single block is encrypted, this part of the encryption algorithm alone is the one-time pad cipher, achieving perfect secrecy. Any chess-related steps afterwards are irrelevant. If new Key 1 is chosen for each plaintext block separately, btw. the paper \cite{K17} can be interpreted both ways (yes and no), this observation extends to the entire ciphertext -- the key is long, the first part of the encryption algorithm is one-time pad and other transformations are redundant for secrecy.

If Key 1 is the same for each block, which is probably the intended construction, a known plaintext attack becomes a problem. It is possible to reconstruct values of Key 1, at least for characters presented in the final position on the chess board, and depending on details of the encryption algorithm even for the entire block.

\subsubsection*{Chess part of the scheme is incomplete}

The scheme encodes moves by creating pairs of squares where a piece was and moved to, respectively. There is no mention whether this encoding is able to correctly work with moves like castling, en passant, and pawn promotion.

\begin{remark}
As a curiosity, the example game used in \cite{K17} is the following one (annotation symbols were added by Stockfish):
\begin{verbatim}
1. b4 e6 2. c3 f5 3. g3 g6 4. Nf3 Bd6?! 5. h4 Nf6 6. Nd4?! a6 7. e3 Bf8 8. Qf3? Nd5? 
9. Nc2 Nc6 10. e4 Ne5 11. Qe2 fxe4 12. d4?? c5?? 13. dxe5 cxb4 14. cxb4?! Rb8?
15. Bg5?! Qc7 16. h5?? Ra8?? 17. hxg6 Be7?! 18. g7 Bxb4+?? 19. Nxb4 h6 20. Bxh6?! d6?!
21. Qh5+ Ke7 22. Qg5+ Kf7 23. gxh8=Q Nxb4 24. Qh7+ Ke8 25. Qh5+ Kd8 26. Bg5+ Qe7
27. Bxe7+ Kd7 28. Bxd6+ Kc6 29. Bxb4 Bd7 30. Qxe4+ Kb6 31. Qhh7 Bc6 32. Qd4#
\end{verbatim}
The game it rather illogical, full of blunders, and white ``overlooks'' multiple mates in 1 opportunities, first one in move 20.
\end{remark}

\subsubsection*{Weak chess games}

Some chess games are only a few moves long, e.g., Scholar's mate, Fool’s mate, and Legal’s Mate. These and other games leave many pieces on their original squares, thus weakening the resulting permutation. The proposal \cite{K17} does not address the possibility of weak chess games for the Chessography, neither how to select suitable chess games for encryption.

\subsubsection*{Chess game permutation is weak}

Let's assume the chess game is generated by a chess engine to be human-like, or selected from a huge pool of human played games. The final composition and placement of pieces is far from statistically random, let alone cryptographically strong random. A simple analysis presented in Section \ref{sec2} demonstrates this convincingly. Using the final position of a chess game and intermediate moves as a permutation component in a cipher is a bad idea.

The claim ``\textsl{The strength of this algorithm is based upon the complexity of the chess game.}'' by the author of Chessography, and then using the estimate for the number of possible chess games to argue the scheme's security, is simply deceiving.

\section{Analysis of chess games final positions}
\label{sec2}

The dataset consists of 100,000 games played on the Lichess server by users with an average rating of 2558. It is a subset of games played in October 2024 \cite{LED}. The dataset contains mostly blitz and rapid games, and excludes bullet time controls. White win rate is 47\%, black win rate is 42\%, and only 11\% of the games are draws. The average length of the game is 43 moves (86 plies).

\begin{figure}[ht]
\centering
\includegraphics[scale=.8]{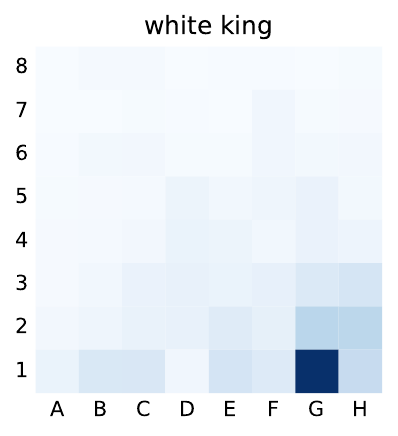}
\hskip4ex
\includegraphics[scale=.8]{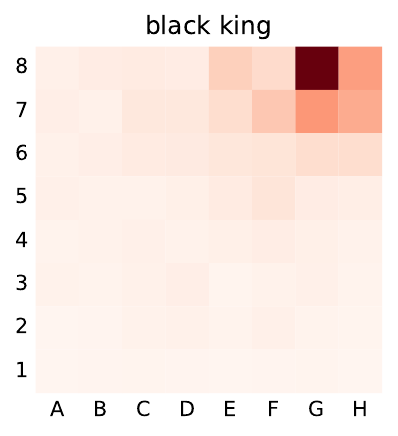}
\caption{Final position of the king}
\label{fig-king}
\end{figure}

Figures \ref{fig-king}--\ref{fig-pawn} on pages \pageref{fig-king}--\pageref{fig-pawn} show heatmaps for the location of different pieces on the chess board for the final position of the game. The heatmaps illustrate the limited randomness of the final positions and subsequently weak (partial) permutations that chess games provide. Colors are scaled individually for each heatmap, therefore the same shade can represents different percentage in distinct heatmaps. Table \ref{tab-max} summarizes the maximal percentages and placement on the board for each piece type.

\begin{table}[ht]
\centering
\begin{tabular}{lrcrc}
piece & \multicolumn{2}{c}{white} & \multicolumn{2}{c}{black} \\
      & max [\%] & square  & max [\%] & square \\
\midrule
king  & 22.28 & g1 & 22.46 & g8 \\
queen &  2.52 & d1 &  3.31 & d8 \\
bishop & 3.83 & g2 &  4.75 & g7 \\
knight & 3.88 & f3 &  4.03 & f6 \\
rook  & 10.45 & a1 & 12.15 & a8 \\
pawn  & 26.22 & f2 & 27.89 & f7
\end{tabular}
\caption{The most common piece placement in the final position}
\label{tab-max}
\end{table}

\subsection*{Conclusion}

This note shows that Chessography is not a good proposal. More importantly, it seems that it is impractical to base a strong encryption scheme on chess games. The rules of chess limit the range of possible moves, and the placement of pieces in final positions is not sufficiently random.

\begin{figure}[ht]
\centering
\includegraphics[scale=.8]{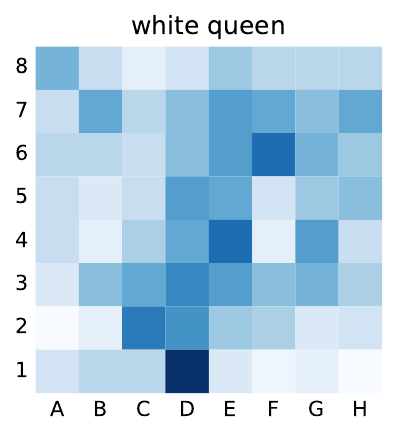}
\hskip4ex
\includegraphics[scale=.8]{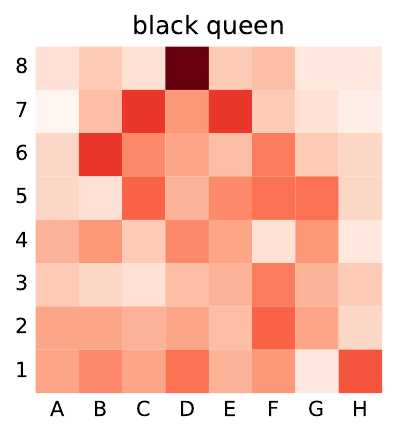}
\caption{Final position of the queen}
\end{figure}

\begin{figure}[ht]
\centering
\includegraphics[scale=.8]{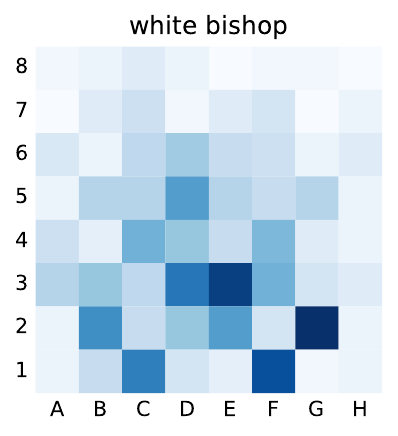}
\hskip4ex
\includegraphics[scale=.8]{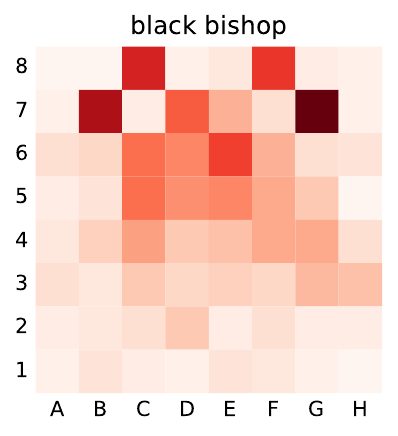}
\caption{Final position of the bishop}
\end{figure}

\begin{figure}[ht]
\centering
\includegraphics[scale=.8]{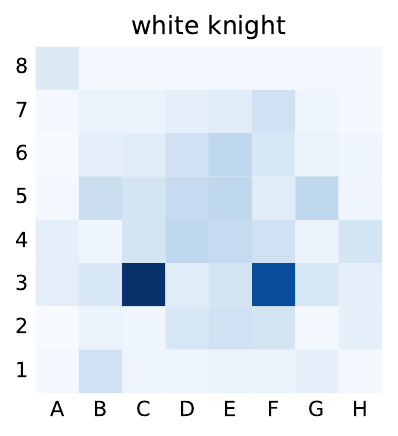}
\hskip4ex
\includegraphics[scale=.8]{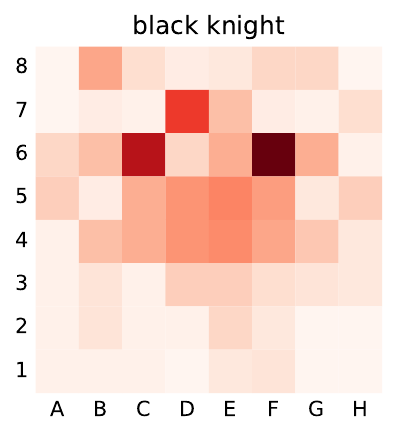}
\caption{Final position of the knight}
\end{figure}

\begin{figure}[ht]
\centering
\includegraphics[scale=.8]{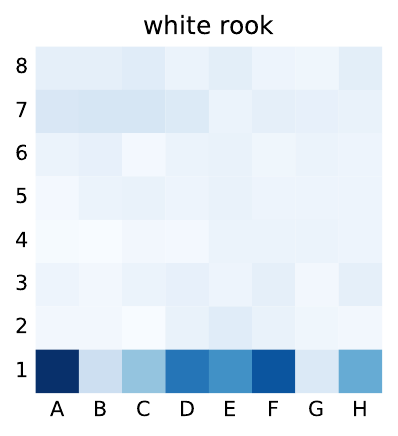}
\hskip4ex
\includegraphics[scale=.8]{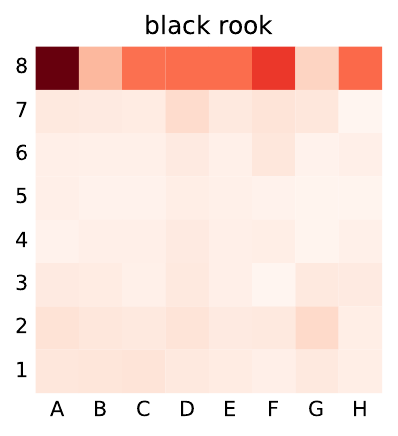}
\caption{Final position of the rook}
\end{figure}

\begin{figure}[ht]
\centering
\includegraphics[scale=.8]{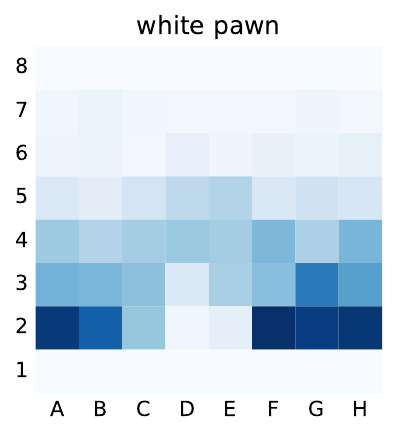}
\hskip4ex
\includegraphics[scale=.8]{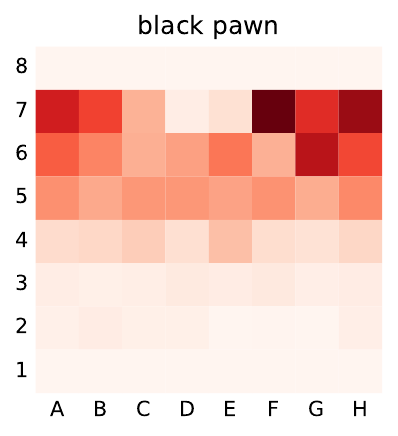}
\caption{Final position of the pawn}
\label{fig-pawn}
\end{figure}

\end{document}